\def\CMP{\sevenrm Commun.\ Math.\ Phys.}
\def\LMP{\sevenrm Lett.\ Math.\ Phys.}
\def\JMP{\sevenrm J.\ Math.\ Phys.}
%
%
\def\today{\number\day .\space\ifcase\month\or
January\or February\or March\or April\or May\or June\or
July\or August\or September\or October\or November\or December\fi, \number \year}
%
%
\newcount \theoremnumber
\def\cleartheoremnumber{\theoremnumber = 0 \relax}

\newcount \subheadlinenumber

\def\SHL #1 {\goodbreak
            \cleartheoremnumber
            \vskip 1cm
            \advance \subheadlinenumber by 1
            {\fourteenrm \noindent {\the\headlinenumber}.{\the\subheadlinenumber}. #1}
            \nobreak \vskip.8cm \rm \noindent}

\def\Prop #1 {
             \advance \theoremnumber by 1
             \vskip .6cm 
             \goodbreak 
             \noindent
             {\bf Propsition {\the\headlinenumber}.{\the\theoremnumber}.}
             {\sl #1}  \goodbreak \vskip.8cm}

\def\Conj#1 {
             \advance \theoremnumber by 1
             \vskip .6cm  
             \goodbreak 
             \noindent
             {\bf Conjecture {\the\headlinenumber}.{\the\theoremnumber}.}
             {\sl #1}  \goodbreak \vskip.8cm} 

\def\Th#1 {
             \advance \theoremnumber by 1
             \vskip .6cm  
             \goodbreak 
             \noindent
             {\bf Theorem {\the\headlinenumber}.{\the\theoremnumber}.}
             {\sl #1}  \goodbreak \vskip.8cm}

\def\Lm#1 {
             \advance \theoremnumber by 1
             \vskip .6cm  
             \goodbreak 
             \noindent
             {\bf Lemma {\the\headlinenumber}.{\the\theoremnumber}.}
             {\sl #1}  \goodbreak \vskip.8cm}

\def\Cor#1 {
             \advance \theoremnumber by 1
             \vskip .6cm  
             \goodbreak 
             \noindent
             {\bf Corollary {\the\headlinenumber}.{\the\theoremnumber}.}
             {\sl #1}  \goodbreak \vskip.8cm} 
%
%
\newcount \equationnumber

\newcount \refnumber

\def\[]    {\global 
            \advance \refnumber by 1
            [{\the\refnumber}]}

\def\# #1  {\global 
            \advance \equationnumber by 1
            $$ #1 \eqno ({\the\equationnumber}) $$ }

\def\% #1 { \global
            \advance \equationnumber by 1
            $$ \displaylines{ #1 \hfill \llap ({\the\equationnumber}) \cr}$$} 

\def\& #1 { \global
            \advance \equationnumber by 1
            $$ \eqalignno{ #1 & ({\the\equationnumber}) \cr}$$}
%
%
\newcount \Refnumber

\def\Ref #1 #2 #3 #4 #5 #6  {\ninerm \global
                             \advance \Refnumber by 1
                             {\ninerm #1,} 
                             {\ninesl #2,} 
                             {\ninerm #3.} 
                             {\ninebf #4,} 
                             {\ninerm #5,} 
                             {\ninerm (#6)}\nobreak} 
\def\Bookk #1 #2 #3 #4       {\ninerm \global
                             \advance \Refnumber by 1
                             {\ninerm #1,}
                             {\ninesl #2,} 
                             {\ninerm #3,} 
                             {(#4)}}
\def\Book{\cr
{\the\Refnumber} &
\Bookk}
\def\Reff{\cr
{\the\Refnumber} &
\Ref}
\def\REF #1 #2 #3 #4 #5 #6 #7   {{\sevenbf [#1]}  & \hskip -9.5cm \vtop {
                                {\sevenrm #2,} 
                                {\sevensl #3,} 
                                {\sevenrm #4} 
                                {\sevenbf #5,} 
                                {\sevenrm #6} 
                                {\sevenrm (#7)}}\cr}
\def\BOOK #1 #2 #3 #4  #5   {{\sevenbf [#1]}  & \hskip -9.5cm \vtop {
                             {\sevenrm #2,}
                             {\sevensl #3,} 
                             {\sevenrm #4,} 
                             {\sevenrm #5.}}\cr}
\def\HEP #1 #2 #3 #4     {{\sevenbf [#1]}  & \hskip -9.5cm \vtop {
                             {\sevenrm #2,}
                             {\sevensl #3,} 
                             {\sevenrm #4.}}\cr}
%
%
\def\bull{$\sqcup \kern -0.645em \sqcap$}
%
%
\def\Def#1{\vskip .3cm \goodbreak \noindent
                                     {\bf Definition.} #1 \goodbreak \vskip.4cm}
\def\Rem#1{\vskip .4cm \goodbreak \noindent
                                     {\it Remark.} #1 \goodbreak \vskip.5cm }

\def\Pr#1{\goodbreak \noindent {\it Proof.} #1 \hfill \bull  \goodbreak \vskip.5cm}

%
%
\def\*{\vskip 1.0cm}      

%
%
\newcount \headlinenumber

\newcount \headlinesubnumber
\def\clearheadlinesubnumber{\headlinesubnumber = 0 \relax}
\def\Hl #1 {\goodbreak
            \cleartheoremnumber
            \clearheadlinesubnumber
            \advance \headlinenumber by 1
            {\bf \noindent {\the\headlinenumber}. #1}
            \nobreak \vskip.4cm \rm \noindent}

\def\SHL #1 {\goodbreak
            \cleartheoremnumber
            \vskip 1cm
            \advance \subheadlinenumber by 1
            {\rm \noindent {\the\headlinenumber}.{\the\subheadlinenumber} #1}
            \nobreak \vskip.4cm \rm \noindent}

\font\twentyrm=cmr17
\font\fourteenrm=cmr10 at 14pt
\font\sevensl=cmsl10 at 7pt
\font\sevenit=cmti7

\font\css=cmss10
\font\Rosch=cmr10 at 9.85pt
\font\Cosch=cmss12 at 9.5pt
\font\rosch=cmr10 at 7.00pt
\font\cosch=cmss12 at 7.00pt
\font\nosch=cmr10 at 7.00pt
%
%
%
%
%
%
%
%
%
%
%
%
%
%
%
%
\def\Z                 {\hbox{{\css Z}  \kern -1.1em {\css Z} \kern -.2em }}
\def\R                 {\hbox{\raise .03ex \hbox{\Rosch I} \kern -.55em {\rm R}}}
\def\N                 {\hbox{\rm I \kern -.55em N}}
\def\C                 {\hbox{\kern .20em \raise .03ex \hbox{\Cosch I} \kern -.80em {\rm C}}}

\def\r                 {\hbox{\raise .03ex \hbox{\rosch I} \kern -.45em \hbox{\rosch R}}}
\def\n                 {\hbox{\hbox{\rosch I} \kern -.45em \hbox{\nosch N}}}
\def\c                 {\hbox{\raise .03ex \hbox{\cosch I} \kern -.70em \hbox{\rosch C}}}

\def\z                 {\hbox{\kern 0.2em {\cal z}  \kern -0.6em {\cal z} \kern -0.3em  }}
\def\1                 {\hbox{\rm \thinspace \thinspace \thinspace \thinspace
                                  \kern -.50em  l \kern -.85em 1}}
\def\unit                 {\hbox{\sevenrm \thinspace \thinspace \thinspace \thinspace
                                  \kern -.50em  l \kern -.85em 1}}
%
%
%
%
%
%
%
%
%
%
%
%

\def\A                 {{\cal A}}

\def\B                 {{\cal B}} 
\def\M                 {{\cal M}}
\def\H                 {{\cal H}} 
 
\def\O                 {{\cal O}}

\def\bra{\langle}
\def\ket{\rangle}

%
%
%
%
%
%
%


\nopagenumbers
\def\Draft  {\hbox{Preprint \today}}
\def\firstheadline{\hss \hfill  \Draft  \hss} 
\headline={
\ifnum\pageno=1 \firstheadline
\else 
\ifodd\pageno \rightheadline 
\else \leftheadline \fi \fi}
\def\rightheadline{\sevenrm CLUSTER ESTIMATES FOR MODULAR STRUCTURES
\hfill \folio } 
\def\leftheadline{\sevenrm \folio \hfill CHRISTIAN D.\ J\"AKEL}
\voffset=2\baselineskip
\magnification=\magstep1
%
%
%
%
%
%
%

\vskip 1cm

\noindent
{\twentyrm Cluster Estimates for Modular Structures}
                  
\vskip 1cm
\noindent
{\sevenrm CHRISTIAN D.\ J\"AKEL\footnote{$^\star$}{\noindent \sevenit  
present address: Dipartimento di Matematica,
via della Ricerca Scientifica, Universit\`a di Roma ``Tor Vergata'', 
I-00133~Roma, Italy, 
E-mail: cjaekel@esi.ac.at} }

\noindent
{\sevenit  Erwin Schr\"odinger Institut, 
Boltzmanng.\ 9, A-1090 Wien, Austria}

\vskip .5cm     
\noindent {\sevenbf Abstract}. {\sevenrm 
The basic ingredients of Tomita-Takesaki modular theory are used 
to establish cluster estimates. Applications to thermal quantum field theory are discussed 
and the convergence of the thermal universal localizing map is proven.}

\vskip 1 cm

%
%
%
%
%

\Hl{Introduction}

\noindent
The cluster theorem of relativistic quantum field theory 
expresses the decay of correlations between clusters of local
observables as the space-like separation between clusters increases.
In $3+1$ dimensions the correlations of the vacuum expectation values of
local observables decrease at least like~$\delta^{-2}$, 
where $\delta$ denotes the space-like distance of the clusters. 
In the presence of a mass-gap one finds an exponential decay 
like~${\rm e}^{-M \delta}$, where~$M$ is the minimal mass in the theory. 
The standard argument (see~[AHR][F]) is based on the spectral properties of the Hamiltonian and 
Einstein causality. Thomas and Wichmann [TW] (see also [D]) provided an alternative 
derivation in case $M > 0$, based on the connection between the representation theory 
of the Poincar\'e group and the modular objects for a wedge-shaped region bounded by two 
characteristic planes.
More recently, the author studied the cluster properties of local observables
in thermal equilibrium. In this case the spectrum of the effective Hamiltonian 
(the generator of time translations in the GNS-representation associated with the
thermal equilibrium state) does not have a mass gap; not even for free massive particles.  
But due to the KMS-condition, which characterizes thermal equilibrium states, there still
exists a tight connection between the spectral properties of the effective Hamiltonian 
and the decay of spatial correlations [J\"a a]. 

As far as observable quantities are concerned the results cited 
sofar are sufficiently general to cover the situations of physical interest.
In principle, one might insist on a representation independent 
description of a physical system in terms of a net 
of (abstract) $C^*$-algebras $\O \to \A(\O)$ and a distinguished 
set of positive linear functionals,
representing the local observables and the experimental preparation possibilities, respectively. 
However, refined mathematical methods for predicting the expectation values of
quantities, which can be compared with experimental data, are currently 
only available, if the observable quantities are represented on some 
Hilbert space~$\H$. Therefore cluster estimates between elements of~$\B(\H)$, 
which do not directly refer to local observables, turn out to be useful. 
For example, the cluster estimates, which will be presented in this letter, provide the clue
for a method which explores the connection between KMS-states 
for different temperatures~[J\"a~b]. 
%
%

\vskip 1cm

\Hl{Main Theorem}

\noindent
As the title indicates, the cluster estimates,
which we will present in this letter, are essentially based on modular theory 
(see e.g.\ [Ta][St][BR][KR]).
However, our starting point is slightly more general. Given a von Neumann algebra 
${\cal N}$ acting on some Hilbert space $\H$,
we will look out for a one-parameter group $U \colon t \mapsto U(t)$,
a conjugate-linear operator $I$ and a vector $\Psi \in \H$
such that for each $N \in {\cal N}$ the function 
\# {t \mapsto U(t) N \Psi }
has an analytic continuation into the strip 
$ {\cal S} (0, 1/2) = \{ z \in \C :
0 \le \Im z \le 1 /2 \}  $  
which satisfies
\# { I U(i /2) N \Psi = N^* \Psi.}
The obvious connection to modular theory is, that if ${\cal N}$ is contained in a
von Neumann algebra ${\cal R}$ with cyclic and separating vector $\Omega$, then 
a possible choice for $U$, $I$ and $\Psi$ are the 
modular group $t \mapsto \Delta^{it}$, the modular conjugation $J$ associated with 
the pair $({\cal R}, \Omega)$ and $\Psi := \Omega$. Given only ${\cal N}$,
there is considerable freedom in the choice of the pair
$({\cal R}, \Omega)$. In fact, we will not assume that ${\cal R}$ is 
explicitly given. Instead we prefer to specify directly the properties of $I$ and $U$.

Once the objects ${\cal N}, U, I,  \Psi$ are specified, we will 
consider the adjoint action of the group $U$ on the von Neumann algebra\footnote{$^\dagger$}
{\sevenrm Recall that Tomita's theorem --- 
the principal result of modular theory --- is that the mapping 
$\scriptstyle  j \colon A \mapsto J  A^* J $
defines a $\scriptstyle *$-anti-isomorph\-ism from $\scriptstyle {\cal R}$ onto 
$\scriptstyle {\cal R}'$.
Given a subalgebra $\scriptstyle {\cal N} \subset {\cal R}$ the 
algebra~$\scriptstyle j({\cal N})$
is called the opposite algebra associated to~$\scriptstyle {\cal N}$.}
generated by ${\cal N}$ and $i ({\cal N}) := Ad \, I {\cal N}$, denoted 
by~${\cal N} \vee  i( {\cal N}) $. 
We will show that, if there exists another von Neumann algebra ${\cal M}$ 
and some~$\delta>0$ such that
\# {  \Bigl[ U(t) \, \bigl({\cal N} \vee i({\cal N}) \bigr) U^{-1}( t) \, , \, {\cal M}
\Bigr] = 0  
\qquad \hbox{for} \quad |t| < \delta,}
or equivalently
\# {  \Bigl[ U(t) \, {\cal N}  U^{-1}( t) \, , \, {\cal M} \vee i({\cal M})
\Bigr] = 0  
\qquad \hbox{for} \quad |t| < \delta,}
then decent spectral properties of the generator of $U$ can be used to bound the expression  
\# { \bigl|  (\Psi\, , \, NM \Psi)  - (\Psi\, , \, N \Psi)
(\Psi\, , \, M \Psi) \bigr| 
}
in terms of inverse powers of the `distance' $\delta$.

\vskip 1cm
We now state our main result.
 
\eject

\Th{Consider two von Neumann algebras ${\cal N}$ 
and ${\cal M}$, both acting on some Hilbert space~$\H$.
Let $I \colon \H \to \H$ be a conjugate-linear isometric operator, satisfying the condition
$I^2 = \1 $ and let $t \mapsto U (t) =: \exp (i H  t)$, $t \in \R$, be a strongly continuous
one-parameter group of unitaries. Assume these objects satisfy the following three conditions:
\vskip .2cm
\halign{ \indent #  \hfil & \vtop { \parindent = 0pt \hsize=34.8em
                            \strut # \strut} 
\cr
(i)    & (Spectral properties).
The generator $H$ of the group $U$
has a unique --- up to a phase --- normalized eigenvector $\Psi \in \H$ for 
the simple eigenvalue $\{ 0 \}$ and otherwise continuous spectrum, not necessarily
semi-bounded. Furthermore, there exist  positive constants $m>0$ and $C > 0$
such that   
\# { \| {\rm e}^{ \mp \lambda H} P^\pm N \Psi \|  
\le C \cdot \lambda^{-m}  \, \| N \Psi \|  \qquad \forall N \in {\cal N},}
where $P^\pm$ denote the projections 
onto the {\sl strictly} positive resp.\ negative spectrum of $H$. 
\cr
(ii)    & (Analyticity properties).
For each $N \in {\cal N}$ the vector valued function
\# { t \mapsto U(t) N \Psi }
admits an analytic continuation into the strip 
$ {\cal S} (0, 1/2) = \{ z \in \C :
0 \le \Im z \le 1 /2 \} $. 
The boundary values are continuous for $\Im z \nearrow 1/2$ and they satisfy
\# { I U(i /2) N \Psi = N^* \Psi.}
Furthermore
\# { I U (z) N \Psi = U (\bar z) I N \Psi  \qquad \forall z \in {\cal S} ( 0, 1/2),
\quad N \in {\cal N}. }
(Note that $\1 \in {\cal N}$, together with $U (z) \Psi = \Psi$ for all 
$z \in {\cal S}_{1/2}$, implies $I  \Psi = \Psi$.) 
\cr
(iii)    & (Distance).
There exists some $\delta > 1$  such that
\# {  \Bigl[ U(t) \, {\cal N}  U^{-1}( t) \, , \, {\cal M} \vee i({\cal M})
\Bigr] = 0  
\qquad \hbox{for} \quad |t| < \delta,}
where $i(T): = I T^*I$ for $T \in \B(\H)$.
\cr
}
\vskip .3cm
\noindent
It follows that
\& { \Bigl|  (\Psi\, , \, NM \Psi) &- (\Psi\, , \, N \Psi)
(\Psi\, , \, M \Psi) \Bigr| 
\cr
& 
\le  \bigl( 5 {\rm e}^m  + 1 \bigr) C
\cdot n_\circ^{-m} \Bigl( \| F N \Psi \| \, \| F M^* \Psi \| 
+ \| F N^* \Psi \| \, \| FM \Psi \| \Bigr), }
where $F := P^+ + P^-$ denotes the projection onto the orthogonal
complement of $\Psi$ and  
$n_\circ := \inf  \{ n \in \N : \delta^{m \over m+1} \le n \le \delta \}$. }
 
\eject

\Rem{For $\delta$ large (compared to $1/2$) the correlations 
decrease like $\sim \delta^{- {m^2 \over m+1}}$. For small~$\delta$  
the finite width of ${\cal S} (0, 1/2)$ is clearly reflected in the
discrete nature of the bounds, which involve a natural number, namely $n_\circ$.}

\noindent
The proof proceeds in several steps:
\vskip .2cm
\noindent
(i) we define a function 
\# { {\cal F}_{M,N} \colon {\cal I}_{\delta} \to \C ,}
bounded and analytic in the infinitely often cut plane  
\# { {\cal I}_{\delta} = \C \backslash \{ z \in \C : 
\Im z = k / 2 , k \in \Z, |\Re z| \ge {\delta}  \}, }
such that 
\# { {\cal F}_{M,N} (0) = \bigl( \Psi \, , \, M  N \Psi ).}
\noindent
(ii) we define a function~$f_{M, N}$, analytic on the twofold cut plane
\# {  {\cal P}_{\delta} = \C \backslash \{ z \in \C: \Im z = 0,  |\Re z | \ge \delta  \} }
such that for arbitrary $n \in \N$:
\& {  {\cal F}_{M , N} (0) - (\Psi \, , \, M \Psi) (\Psi \, , \, N \Psi ) &=
\sum_{ l= -n}^{n} f_{ M , N} (i l )  
+ \sum_{ l= -n}^{ n-1 } f_{ I M^*I , N} \bigl(i (l + 1/2) \bigr)  
\cr
& \qquad + \bigl( U (-in) P^- N^* \Psi \, , \,   M \Psi \bigr)
\cr
& \qquad + \bigl(\Psi \, , \, M  U (-in) P^-  N  \Psi \bigr) .}
\noindent
(iii) we derive bounds for the terms on the r.h.s.\ of (16).
\vskip .2cm
\noindent
(iv) we optimize the natural number $n \in \N$ in (16).

\vskip 1cm

\Pr{(i)  Let $M \in {\cal M} \vee i ({\cal M})$ and $N \in {\cal N}$. Then
\& {\lim_{\Im z \nearrow 1/2}  \bigl( \Psi \, , \, M U (z) N \Psi \bigr)
& =  \bigl( M^* I \Psi \, , \, U ( \Re z ) U (i /2 ) N \Psi \bigr)
\cr
& =  \bigl( U (\Re z) I  U (i /2 ) N \Psi  \, , \,I M^*I \Psi \bigr)
\cr
&  = \bigl( U (\Re z) N^* \Psi \, , \,I M^*I \Psi \bigr)
\cr
&  = \bigl( \Psi \, , \, U (\Re z) N  U (-\Re z)I M^*I \Psi \bigr). }
From (10) and the fact that ${\cal M} \vee i ({\cal M})$
is invariant under the adjoint action of $I$, it follows that
\# { [ I M^*I \, , \, U (t) N U (-t) ] = 0  \qquad \hbox{for} \quad
   | t | < \delta .  }
\eject
\noindent
Thus, for $| \Re z | < \delta $, 
\& {\lim_{\Im z \nearrow 1/2}  \bigl( \Psi \, , \, M U (z) N \Psi \bigr)
&= \bigl( \Psi \, , \,I M^*I U (\Re z) N \Psi \bigr)
\cr
&= \lim_{\Im z \searrow 1/2} \bigl( \Psi \, , \,I M^*I U (z -i/2) 
N \Psi \bigr) .}
Using the Edge-of-the-Wedge Theorem [SW]
we conclude that there exists a function 
\# { {\cal F}_{M,N} \colon {\cal I}_{\delta} \to \C ,}
bounded and analytic in the infinitely often cut plane  
\# { {\cal I}_{\delta} = \C \backslash \{ z \in \C : 
\Im z = k / 2 , k \in \Z, |\Re z| \ge {\delta}  \} }
with the following properties:
${\cal F}_{M,N}$ is periodic in $\Im z$ with period $1$  
and, for $-1 \le \Im z < 1$, ${\cal F}_{M,N}$ is given by 
\# {
{\cal F}_{M,N} (z) 
= 
\left\{
\eqalign{
&  { \bigl( \Psi \, , \, I M^*I  U (z - i/2)  N  \Psi \bigr), }
\cr
&  { \bigl( \Psi \, , \, M U (z) N \Psi ), }
\cr
&  { \bigl( \Psi \, , \, I M^*I  U (z + i/2)  N  \Psi \bigr), }
\cr
&  { \bigl( \Psi \, , \, M U (z + i) N \Psi \bigr), }}
\right\} 
{\rm \ for \ }  
\left\{
\eqalign{
& { 1 /2  \le \Im z  <  1,}  
\cr
& { 0  \le \Im z   <  1 / 2 ,}  
\cr
& { - 1 /2  \le \Im z  <  0 ,}  
\cr
& { - 1   \le \Im z  <  - 1 / 2.}}  
\right\} 
}
(ii) Now let $P^+$, $P^-$ and $P^{ \{ 0 \} } = | \Psi \ket \bra \Psi |$ denote the spectral 
projections  onto the strictly positive, the strictly negative and the discrete 
spectrum  $\{ 0 \} $ of~$H$.  
The function $f^+_{M, N} \colon \{ z \in \C : \Im z \ge 0 \} \to \C$, 
\# { z \mapsto
\bigl( \Psi \, , \, M  U (z) P^+ N \Psi \bigr)
- \bigl( \Psi \, , \, N U (- z) P^-  M \Psi \bigr), }
is analytic in the upper half plane $\Im z > 0$, while the function  
$f^-_{M, N} \colon \{ z \in \C : \Im z \le 0 \}  \to \C$,  
\# {  z \mapsto
- \bigl( \Psi \, , \, M U (z) P^- N \Psi \bigr) 
+ \bigl( \Psi \, , \, N U (-z) P^+ M \Psi \bigr),}
is analytic in the lower half plane $\Im z < 0$. Both functions have continuous 
boundary values for $\Im z \searrow 0$ and $\Im z \nearrow 0$, respectively.
Thus  
\# {   (\Psi \, , \,   [ M  \, , \, U (t) N U (-t) ] \Psi) = 
f^+_{ M, N} (t) - f^-_{ M, N } (t)  
\qquad \forall t \in \R .}
Because of the commutator on the l.h.s., the discrete spectral value $\{ 0 \}$ 
does not contribute to the r.h.s. By assumption, the l.h.s.\ vanishes for $|t| < \delta$.
Consequently, the boundary values for $\Im z \searrow 0$ and $\Im z \nearrow 0$, 
respectively, of the functions defined in (23) and (24) coincide for $| \Re z| < \delta $.
We conclude that there exists a 
function~$f_{M, N}$, analytic on the twofold cut plane
$  {\cal P}_{\delta} = \C \backslash \{ z \in \C: \Im z = 0,  |\Re z | \ge \delta  \} $
such that
\# {
f_{M, N} (z) = 
\left\{
\eqalign{
&  {f^+_{M, N} (z)}
\cr
&
  {f^-_{M, N} (z) }
}
\right\} 
{\rm \ for \ }  
\left\{
\eqalign{
&  {\Im z > 0,}
\cr
&
  {\Im z < 0.}
}
\right\} 
}
By definition,
\# {  \bigl( \Psi \, , \, M U (z) P^+ N \Psi \bigr) 
- f^+_{M, N} (z)   =  \bigl( N^* \Psi \, , \,  U (-z) P^-  M \Psi \bigr) 
\qquad \forall | \Im z | \ge 0.}
The r.h.s.\ equals
\& { \bigl( I U (- z) P^-  M \Psi \, , \,I N^* \Psi \bigr)
& = \bigl( U (- \bar z) P^+I M \Psi \, , \, U (i / 2) N \Psi\bigr)
\cr
& = \bigl( \Psi \, , \,I M^*I U (z + i/2)  P^+ N \Psi \bigr) 
\qquad \forall | \Im z | \ge 0.}
Thus
\# {  \bigl( \Psi \, , \, M U (z) P^+ N \Psi \bigr) 
=  f^+_{M, N} (z) + \bigl( \Psi \, , \,I M^*I U (z + i/2)  P^+ N \Psi \bigr) 
\qquad \forall | \Im z | \ge 0.}
Since $I M^* I \in {\cal M} \vee i({\cal M})$ for $M \in {\cal M} \vee i({\cal M})$,
we can now iterate this identity:
\& {  \bigl(\Psi \, , \, M U (t) P^+ N \Psi \bigr) 
& = f_{M, N} (t) + f_{I M^*I, N} ( t + i/2) +
\bigl( \Psi \, , \, M U (t+i) P^+ N \Psi \bigr)
\cr
& = \sum_{l = 0}^{n} f_{M, N} (t + il) + 
\sum_{l = 0}^{n-1} f_{I M^*I, N} \bigl( t + i(l +1/2) \bigr) 
\cr
& \qquad +
\bigl( U (-in) P^- N^* \Psi \, , \,  U (-t)   M \Psi \bigr). }
On the other hand, for $l + 1/2 < 0$,
\& {  \bigl( \Psi \, , \, M U (t+il)  P^- N \Psi \bigr) & + f_{M, N} (t + il)
=   
\cr
& = 
\bigl( U (i/2) N^* \Psi \, , \,  
U \bigl( - t - i(l + 1 /2) \bigr)  P^+  M \Psi \bigr)
\cr
& = \bigl(I U \bigl( - t - i(l + 1 /2) \bigr)  P^+  M \Psi \, , \,  N \Psi \bigr)
\cr
& = \bigl( U \bigl( - t + i(l + 1/2) \bigr)  
P^- I M^*I \Psi \, , \,  N \Psi \bigr)
\cr
& = \bigl( \Psi \, , \,I M^*I U \bigl( t + i(l + 1/2) \bigr)  
P^- N \Psi \bigr) .}
By iteration 
\& {  \bigl( \Psi \, , \, M U (in) P^- U (t) N \Psi \bigr) 
& + \sum_{l = -n}^{l= -1} f_{M, N} (t + il) 
+ \sum_{l = -n}^{l = -1} f_{I M^*I, N} \bigl( t + i(l +1/2) \bigr) 
= \cr
& \qquad = \bigl( \Psi \, , \, M U (t) P^- N \Psi \bigr)  \qquad 
\forall n \in \N.}
The following identity, which holds for arbitrary $n \in \N$ and 
for $|t| < \delta$, now follows by addition:
\& {  {\cal F}_{M , N} (t) - (\Psi \, , \, M \Psi) (\Psi \, , \, N \Psi ) &=
\sum_{ l= -n}^{n} f_{ M , N} (t + i l )  
+ \sum_{ l= -n}^{ n-1 } f_{ I M^*I , N} \bigl(t + i (l + 1/2) \bigr)  
\cr
& \qquad + \bigl( U (-in) P^- N^* \Psi \, , \,  U (-t)   M \Psi \bigr)
\cr
& \qquad + \bigl(\Psi \, , \, M  U (-in) P^- U (t) N  \Psi \bigr).}
\noindent
(iii) The assumptions on the spectral properties of $H$ allow us to derive bounds on the 
terms on the r.h.s.\ of (33). For 
$\lambda > 0$  
\& {  \sup_{t \in \r} |f_{M,N} (t + i \lambda) |
& \le \sup_{t \in \r}
\Bigl|  \bigl( \Psi \, , \, M  U \bigl( - i (t + i \lambda) \bigr) P^{+} 
N \Psi \bigr) \Bigr|
\cr
& \qquad + \sup_{t \in \r}
\Bigl|  \bigl( U \bigl(- i (t - i \lambda) \bigr) P^{-} N^* \Psi \, , \,   
M \Psi \bigl) \Bigr|
\cr
&
\le  C  \cdot  \lambda^{-m}  \Bigl( \| M^* \Psi \| \,  \| N \Psi \| 
+ \| M \Psi \| \,  \| N^* \Psi \| \Bigr).} 
Similar bounds hold for $\lambda < 0$. 
Jensen's inequality can be used to derive the following uniform bounds 
(see [J\"a a], Lemma 2.2):
\# { | f_{M,N}( i r) |  \le C   
\cdot  \Bigl( { {\rm e} \over \delta } \Bigr)^{m} 
\Bigl( \| M^* \Psi \| \,  \| N \Psi \| 
+ \| M \Psi \| \,  \| N^* \Psi \| \Bigr)  
\qquad  \hbox{\rm for} \quad 0 \le r \le \delta .}
Recall that $\| I M^* I \| = \| M \| $ and, hence,
\& {  \Bigl| (\Psi \, , \, \bigl( M - (\Psi \, , \,  M  \Psi ) \bigr)  
N \Psi ) \Bigr|  
&=
\Bigl| \sum_{ l= -n}^{n} f_{M , N} ( i l )  \Bigr|
+ \Bigl| \sum_{ l= -n}^{ n-1 } f_{ I M^* I , N} \bigl( i (l + 1/2) \bigr) \Bigr| 
\cr
&  \quad + \Bigl(  \| M \Psi \| \, \| U(-in) P^- N^*  \Psi \| + 
\| M^* \Psi \| \, \| U(-in) P^- N  \Psi \| \Bigr)
\cr
& \le
\inf_{n \le \delta } 
\Bigl[ (4n+1) \Bigl( { \delta \over {\rm e} } \Bigr)^{-m } +
n^{-m}   \Bigr] \times
\cr
& \qquad \qquad \times C   \Bigl( \| M^* \Psi \| \,  \| N \Psi \| 
+ \| M \Psi \| \,  \| N^* \Psi \| \Bigr).
}
\vskip .2cm
\noindent
(iv) Set 
$n_\circ := \inf  \{ n \in \N : \delta^{m \over m+1} \le n \le \delta \} $. It follows that 
\# {  \inf_{n \le \delta } 
\Bigl[ (4n+1) \Bigl( { \delta \over {\rm e} } \Bigr)^{-m } +
n^{-m}   \Bigr]  
\le  \bigl( 5{\rm e}^m  + 1 \bigr)   
\cdot n_\circ^{-m} . }
\vskip .2cm
\noindent
(v) As noted by Thomas and Wichmann [TW], one can obtain sharper bounds by inserting 
the self-adjoint projection $F$ onto the orthogonal complement of $\Psi$.
This can be achieved here by replacing $M$ by $M - (\Psi \, , \, M \Psi)$ and 
$N$ by $N - (\Psi \, , \, N \Psi)$ in (36).}

The next theorem concernes certain uniform bounds on the correlations.

\Th{Consider two von Neumann algebras ${\cal N}$ 
and ${\cal M}$, both acting on some Hilbert space~$\H$.
Let $I \colon \H \to \H$ be a conjugate-linear isometric operator, satisfying the condition
$I^2 = \1 $ and let $t \mapsto U (t) := \exp(i H  t)$, $t \in \R$, be a strongly continuous
one-parameter group of unitaries. Assume these objects satisfy the following three conditions:
\vskip .2cm
\halign{ \indent #  \hfil & \vtop { \parindent = 0pt \hsize=34.8em
                            \strut # \strut} 
\cr
(i)    & (Spectral properties).
The generator $H$ of the group $U$
has a unique --- up to a phase --- normalized eigenvector $\Psi \in \H$ for 
the simple eigenvalue $\{ 0 \}$ and otherwise continuous spectrum, not necessarily
semi-bounded. The maps $\Theta^\pm_{\lambda} \colon {\cal N} \to \H $  
\# {  N   \mapsto {\rm e}^{\mp \lambda H } P^\pm N \Psi ,
\qquad \lambda > 0 ,} 
where $P^\pm$ denote the projections 
onto the {\sl strictly} positive resp.\ negative spectrum of $H$, 
are nuclear and there exist positive constants
$C$ and $m$ such that the nuclear norm $\| \, . \, \|_1$ of $ \Theta^\pm_{\lambda} $ 
satisfies
\# { \|   \Theta^\pm_{\lambda}\|_1 
\le C \cdot \lambda^{-m}  .}
\cr
(ii)    & (Analyticity properties).
For each $N \in {\cal N}$ the vector valued function
\# { t \mapsto U(t) N \Psi }
admits an analytic continuation into the strip 
$ {\cal S} (0, 1/2) = \{ z \in \C :
0 \le \Im z \le 1 /2 \} $. 
The boundary values are continuous for $\Im z \nearrow 1/2$ and satisfy
\# { I U(i /2) N \Psi = N^* \Psi.}
Furthermore
\# { I U (z) N \Psi = U (\bar z) I N \Psi  \qquad \forall z \in {\cal S} ( 0, 1/2),
\quad N \in {\cal N}. }
\cr
(iii)    & (Distance).
There exists some $\delta > 1$  such that
\# {  \Bigl[  U(t) \,  {\cal N}  U^{-1}(t) \, , \, {\cal M}  \vee i({\cal M}) \Bigr] = 0  
\qquad \hbox{for} \quad   |t| < \delta ,}
where $i(T): = I T^*I$ for $T \in \B(\H)$.
\cr
}
\vskip .3cm
\noindent
It follows that for two arbitrary families of operators $M_j \in {\cal M}$ and
$N_j \in {\cal N}$, $j \in \{  1 , \ldots , j_\circ \}$  
\& {\Bigl| \Bigl( \Psi \, , \, \sum_{j=1}^{j_\circ} M_j N_j \Bigr)
&-  \sum_{j=1}^{j_\circ} \bigl( \Psi \, , \, M_j \Psi \bigr) 
\bigl( \Psi \, , \, N_j \Psi \bigr) \Bigr|  
\le
\bigl( 8 {\rm e}^m  + 2 \bigr)C 
\cdot n_\circ^{-m}
\Bigl\| \sum_{j=1}^{j_\circ} M_j N_j \Bigr\| . }
where  $n_\circ := \inf  \{ n \in \N : \delta^{m \over m+1} \le n \le \delta \}$. }

\eject

\Pr{The identity (33) generalizes to
\& { \Bigl| \bigl( \Psi   \, , \, \sum_{j=1}^{j_\circ} \bigl( M_j 
- (\Psi \, , \, M_j \Psi) \1 \bigr) N_j \Psi\bigr) 
\Bigr| 
\le &
\sum_{ l= -n }^{n-1} \Biggl[ \Bigl| \sum_{j=1}^{j_\circ}  f_{M_j,N_j} (i l ) \Bigr| 
+ \Bigl| \sum_{j=1}^{j_\circ}  f_{ I M_j^*I , N } \bigl( i (l + 1/2) \bigr) \Bigr| \Biggl]
\cr
& 
+
\Bigl| \sum_{j=1}^{j_\circ}
\bigl( \Psi \, , \, M_j  \bigl( U (in) P^+ 
+ U (-in) P^-  \bigr)  N_j \Psi \bigr) \Bigr| .}
Introducing sequences of vectors 
$ \Phi^{(\lambda)}_k \in \H $ and of linear 
functionals $\phi^{(\lambda)}_k \in {\cal N}^*$ 
such that 
\# { \Theta^+_{\lambda } (N_j) = \sum_k \phi^{(\lambda)}_k (N_j) \cdot
\Phi^{(\lambda)}_k }
one finds
\& { \Biggl| \sum_{j=1}^{j_\circ}
\bigl( M^*_j  
\Psi \, ,  \,   U (in)  P^{+} N_j   \Psi \bigr) \Biggr|  
&= \Biggl|  \sum_{k=1}^\infty \sum_{j=1}^{j_\circ}
\phi^{(n)}_k (N_j) \cdot    \Bigl( M^*_j  
\Psi \, , \,  \Phi^{(n)}_k \Bigr)    \Biggr| 
\cr
&
\le 
\sum_{k} \| \phi^{(n)}_k  \| \, \|  \Phi^{(n)}_k \|  
\cdot
\Bigl\| \sum_{j=1}^{j_\circ} M_j N_j \Bigr\|
\cr
&
\le    C   \cdot   n^{-m}   
\Bigl\| \sum_{j=1}^{j_\circ} M_j N_j \Bigr\| .}
A similar bound holds for the term containing $P^-$ in (45). The
same method can be used to show
\# {
\left\{
\eqalign{
&  { \sup_{t \in \r}  \Bigl| \sum_{j=1}^{j_\circ}  f_{M_j , N_j} (t \pm i \lambda) \Bigr|  }
\cr
&  { \sup_{t \in \r}  \Bigl| \sum_{j=1}^{j_\circ}  f_{I M^*_jI, N_j} (t \pm i \lambda) \Bigr| }
}
\right\} 
\le  2  C  
\cdot  | \lambda |^{- m }
\Bigl\| \sum_{j=1}^{j_\circ} M_j N_j \Bigr\|  .   }
Hence
\& {  \Bigl| \bigl( \Psi   \, , \, \sum_{j=1}^{j_\circ} \bigl( M_j 
- (\Psi \, , \, M_j \Psi) \1 \bigr) N_j \Psi\bigr) 
\Bigr| 
& \le 
\Bigl| \sum_{ l= -n}^{n-1} \sum_{j=1}^{j_\circ} f_{M_j , N_j} ( i l )  \Bigr|
+ \Bigl| \sum_{ l= -n}^{ n-1 } \sum_{j=1}^{j_\circ} 
f_{I M_j^* I , N_j} \bigl( i (l + 1/2) \bigr) \Bigr| 
\cr
& \quad
+
\Bigl| \sum_{j=1}^{j_\circ}
\bigl( \Psi \, , \, M_j  \bigl( U (in) P^+ 
+ U (-in) P^-  \bigr)  N_j \Psi \bigr) \Bigr|  
\cr
&  \le
\inf_{n \le \delta  } 
\Bigl( 4n \Bigl( { \delta \over {\rm e} } \Bigr)^{-m } +
n^{-m}   \Bigr) \cdot 2C    \cdot
\Bigl\| \sum_{j=1}^{j_\circ} M_j N_j \Bigr\|. }
Once again, $n_\circ := \inf  \{ n \in \N : \delta^{m \over m+1} \le n \le \delta \}$
provides a convenient choice for $n \in \N$ in~(49).  }


\Hl{Applications to Thermal Field Theories} 

\noindent
We start from a local quantum field theory $\O \to \A(\O)$, specified by a net 
of (abstract) $C^*$-algebras~$\A(\O)$ and a one parameter group of 
automorphsims $\tau$, representing the time evolution
(as described in the monograph by Haag [H]). 
The Hermitian elements of ${\cal A} (\O)$ are interpreted as the 
observables which can be measured at times and locations in $\O$.
The time evolution acts geometrically, i.e.,
\# { \tau_t \bigl( \A(\O) \bigr) = \A(\O + te) \qquad \forall t \in \R,}
where $e$ is a unit vector denoting the time direction with respect to a given Lorentz-frame.
The net $\O \to {\cal A}(\O)$, for mathematical
convenience embedded in the $C^*$-algebra
\# {   \A 
= \overline {\bigcup_{\O \subset \r^4} \A (\O) }^{C^*} ,}
is subject to a number of physically significant 
conditions which play no role for the present result except for the 
principle of locality:
If two local regions $\O_1$ and $\O_2$ are space-like separated by some 
distance $\delta$,
then the corresponding algebras commute, i.e.,
\# { {\cal A}(\O_2) \subset {\cal A}^c (\O_1)  ,} 
where ${\cal A}^c (\O_1) := \{ a \in \A : [a,b]=0 \, \,  
\forall b \in {\cal A}(\O_1) \}$.

\vskip .5cm
Thermal equilibrium states are characterized by the KMS-condition [HHW].
Given a KMS-state $\omega_\beta$, the GNS-representation 
$(\pi_\beta, \Omega_\beta, \H_\beta)$ gives rise to a
thermal field theory, specified by a net 
of von Neumann algebras
\# { \O \to {\cal R}_\beta(\O) := \pi_\beta \bigl(\A(\O)\bigr)'' .}
Due to the time invariance of $\omega_\beta$,
the time evolution can be unitarily implemented in the representation $\pi_\beta$.
It coincides --- up to rescaling --- with the 
modular group 
\# {t \mapsto \Delta_\beta^{it} }
associated with the GNS-vector $\Omega_\beta$ and the
von Neumann algebra ${\cal R}_\beta := \pi_\beta ( \A )''$.

\SHL{A First Example} 

\noindent
Let $\O$ be some open and bounded space-time region. Since  ${\cal R}_\beta (\O)
\subset {\cal R}_\beta$, condition~(ii) of Theorem 2.1 is automatically satisfied, if 
$\exp ( - \beta H)$ and $I$ are identified with  
the modular objects $(\Delta_\beta, J_\beta)$ associated with the 
pair $({\cal R}_\beta,\Omega_\beta)$ and $\Psi := \Omega_\beta$.
The action of the modular group on the 
algebra 
\# { {\cal M}_\beta (\O) := {\cal R}_\beta(\O) \vee j_\beta \bigl( {\cal R}_\beta(\O) \bigr), }
where $j_\beta (T) := J_\beta T^* J_\beta$ for all $T \in \B(\H_\beta)$, is geometrical, i.e.,  
\# { \Delta_\beta^{it} {\cal M}_\beta (\O) \Delta_\beta^{-it} =
{\cal M}_\beta (\O + t \beta e) \qquad \forall t \in \R.}
In fact, the map $\O \to {\cal M}_\beta (\O)$ preserves inclusions and
respects the local structure, i.e.,
\# { {\cal M}_\beta (\O_1) \subset  {\cal M}_\beta (\O_2)' \qquad \hbox{if} 
\quad \O_1 \subset \O_2' .} 
If the effective Hamiltonian $H$ --- the generator of the modular group ---
has decent infrared properties (as required in condition (i)), then
Theorem 2.1 provides cluster estimates between elements in 
\# { {\cal N} := {\cal R}_\beta(\O)
\qquad \hbox{and} \qquad {\cal M} := \Bigl( \bigvee_{ |t| < \delta } 
{\cal M}_\beta (\O + t \beta e) \Bigr)'  }
in the vector state induced by $\Omega_\beta$.
Note that if $(\O + t \beta e) \subset \hat{\O}$ for all $|t| < \delta$, then
${\cal M}_\beta (\hat{\O}') \subset {\cal M}$. 

The reader may have noticed that condition (ii) of Theorem 2.1 fails, if we try to
set ${\cal N}:=  {\cal M}_\beta (\O)$, $U(t) := \Delta_\beta^{it}$ and $\Psi := \Omega_\beta$:
for an element $A \in {\cal R}_\beta (\O)$ the function
\# {t \mapsto \Delta_\beta^{it} A \Omega_\beta, \qquad t \in \R ,} 
allows an analytic continuation into the strip $S(0, 1/2)$, whereas
for an element $B \in j_\beta \bigl({\cal R}_\beta (\O) \bigr)$ the function
\# {t \mapsto \Delta_\beta^{it} B \Omega_\beta, \qquad t \in \R ,} 
allows an analytic continuation into the strip $S(-1/2, 0)$. For an  
arbitrary element $N \in {\cal M}_\beta (\O)$ the function
\# {t \mapsto \Delta_\beta^{it} N \Omega_\beta, \qquad t \in \R ,} 
will in general not allow any analytic continuation. This is why 
we had to restrict 
ourselves in this example to the case ${\cal N} \subset {\cal R}_\beta$.

\Rem{Previous
cluster estimates [J\"a a] only covered the case when both
${\cal N}$ and ${\cal M}$ where subalgebras of~${\cal R}_\beta$.
Although the description 
of a physical system should not depend on 
non-observable, auxiliary quantities, it is interesting to note
that the spatial correlations between elements in ${\cal R}_\beta (\O)$ and
$j_\beta \bigl( {\cal R}_\beta (\hat{\O}') \bigr)$
in the vector state induced by the KMS-vector $\Omega_\beta$ decrease as the 
space-like distance between $\O$ and 
$\hat{\O}'$ increases. This suggests a geometric interpretation of the 
map $\O \to j_\beta \bigl( {\cal R}_\beta (\O) \bigr)$; it is not just an abstract labeling.}

\SHL{A Refined Example}

\noindent
If the net $\O \to {\cal R}_\beta (\O)$ satisfies the split property, then we can
improve the previous result: we will show that Theorem 2.2 allows us to estimate the correlations 
between elements in 
\# { {\cal N}:=  {\cal M}_\beta (\O) \qquad \hbox{and}
\qquad
{\cal M} := {\cal M}_\beta (\hat{\O})'}
in the vector state induced by the KMS-vector $\Omega_\beta$, as the space-like distance 
between $\O$ and $\hat{\O}'$ increases. The proof of this result is involved and
we proceed in several steps.  

\vskip .5cm
\noindent
i.) We will invesitgate the following geometrical situation: we consider four space-time regions
$\O,\O_1,\O_2$ and $\hat{\O}$ such that
\# {\O \subset \subset \O_1 + t \beta \cdot e \subset \O_2 \subset \subset \hat{\O} 
\qquad \forall |t| < \delta, \quad \delta > 0.}
{\it Notation}.
By $\O \subset \subset \O_1$ --- note that $t=0$ was not excluded in (63) ---
we mean that the closure of the open and bounded space-time region
$\O$ lies in the interior of the open space-time region $\O_1$. We will also use
the abreviations $\Gamma := (\O, \O_1)$ and $\Lambda := (\O_2, \hat{\O})$.
\vskip .5cm
\noindent
ii.) The split property for the net $\O \to {\cal R}_\beta (\O)$  
provides us with two type I factors 
${\cal S}_\Gamma$ and ${\cal S}_\Lambda$ such that
\# {{\cal R}_\beta (\O) \subset {\cal S}_\Gamma 
\subset {\cal R}_\beta (\O_1 + t e) \subset 
{\cal R}_\beta (\O_2) \subset {\cal S}_\Lambda \subset {\cal R}_\beta (\hat{\O}) 
\qquad \forall |t| < \delta.}
\vskip .5cm
\noindent
iii.) The existence of the type I factors ${\cal S}_\Gamma$ and ${\cal S}_\Lambda$ implies 
(see [J\"a d]) that 
there exist two vectors $\Omega_\Gamma$ and $\Omega_\Lambda \in \H_\beta$ in the natural 
postive cones
\# { {\cal P}^\natural \bigl( {\cal R}_\beta (\O) \vee {\cal R}_\beta (\O_1)' , 
\Omega_\beta \bigr) \quad \hbox{and} \quad 
{\cal P}^\natural \bigl( {\cal R}_\beta (\O_2) \vee {\cal R}_\beta (\hat{\O})' , 
\Omega_\beta \bigr), }
respectively, such that
\# { (\Omega_\Gamma \, , \, A B \Omega_\Gamma) =  (\Omega_\beta \, , \, A \Omega_\beta) 
(\Omega_\beta \, , \, B \Omega_\beta)}
for all $A \in {\cal R}_\beta (\O)$ and all $B \in {\cal R}_\beta (\O_1)'$ and
\# { (\Omega_\Lambda \, , \, CD   \Omega_\Lambda) =  (\Omega_\beta \, , \, C \Omega_\beta) 
(\Omega_\beta \, , \, D \Omega_\beta)}
for all $C \in {\cal R}_\beta (\O_2)$ and all $D \in {\cal R}_\beta (\hat{\O})'$.
$\Omega_\Gamma$ is cyclic and separating for 
${\cal R}_\beta (\O) \vee {\cal R}_\beta (\O_1)'$ and 
$\Omega_\Lambda$ is cyclic and separating for
${\cal R}_\beta (\O_2) \vee {\cal R}_\beta (\hat{\O})'$.
Because of the Reeh-Schlieder property 
of $\Omega_\beta$ (see [J\"a e]), the product vectors $\Omega_\Gamma$ and $\Omega_\Lambda$ 
are cyclic even for~${\cal M}_\beta (\O)$.  
\vskip .5cm
\noindent
iv.) The existence of $\Omega_\Lambda$ implies that the 
von Neumann algebra generated by ${\cal R}_\beta (\O_2) $ 
and $ j_\beta \bigl( {\cal R}_\beta (\O_2) \bigr) \subset 
{\cal R}_\beta (\hat{\O})$, namely
\# {  {\cal M}_\beta (\O_2) := {\cal R}_\beta (\O_2) 
\vee j_\beta \bigl( {\cal R}_\beta (\O_2) \bigr) , }
is naturally isomorphic to the $W^*$-tensor product of
${\cal R}_\beta (\O_2) $ and $ j_\beta \bigl( {\cal R}_\beta (\O_2) \bigr)$:
\# {  {\cal M}_\beta (\O_2) \cong {\cal R}_\beta (\O_2) 
\otimes j_\beta \bigl( {\cal R}_\beta (\O_2) \bigr).  }
\vskip .5cm
\noindent
v.) For $A \in {\cal R}_\beta (\O_2) $ and 
$ B \in j_\beta \bigl( {\cal R}_\beta (\O_2) \bigr)$ the function
\# {t \mapsto \Delta_\beta^{it} A \Omega_\beta \otimes \Delta_\beta^{-it} B \Omega_\beta,
\qquad t \in \R ,} 
allows an analytic continuation into the strip $S(0, 1/2)$. 
\vskip .5cm
\noindent
vi.) Since $j_\beta \bigl( {\cal R}_\beta (\O_2) \bigr) \subset {\cal R}_\beta (\hat{\O})'$,
the product vector $\Omega_\Lambda$ can  be utilized to specify the isomorphism (69):
The unitary operator 
$V_\Lambda \colon \H_\beta 
\to \H_\beta  \otimes \H_\beta$ defined by linear extension of
\# { V_\Lambda CD   \Omega_\Lambda 
=  C \Omega_\beta \otimes D \Omega_\beta ,}
where $C \in {\cal R}_\beta (\O_2) $ and 
$D \in {\cal R}_\beta (\hat{\O})'$, satisfies   
\# {V_\Lambda {\cal R}_\beta (\O_2) V_\Lambda^* = {\cal R}_\beta (\O_2) \otimes \1 \qquad
\hbox{and} \qquad 
V_\Lambda {\cal R}_\beta (\hat{\O})' V_\Lambda^* = \1 \otimes {\cal R}_\beta (\hat{\O})'  }
and maps the type I factor ${\cal S}_\Lambda$ onto $\B(\H_\beta) \otimes \1 $.
We emphasize that our specification of the isomorphism (69) depends on the choice of both
$\O_2$ and $\hat{\O}$. 
\vskip .5cm
\noindent
vii.) Given the isometry $V_\Lambda$ specified in (71), 
a one-parameter group of unitaries 
$t \mapsto \Delta_\Lambda^{-it} \colon \H_\beta \to \H_\beta$  
and an anti-unitary operator $J_\Lambda \colon \H_\beta \to \H_\beta$ are specified
by linear extension of
\# {  \Delta_\Lambda^{-it} CD \Omega_\Lambda  :=  V_\Lambda^*  \Bigl(  
 \Delta_\beta^{-it} C \Omega_\beta \otimes \Delta_\beta^{it} D \Omega_\beta \Bigr), 
\qquad t \in \R,}
and, respectively, 
\# {  J_\Lambda  CD \Omega_\Lambda  :=  V_\Lambda^*  \Bigl(  
 J_\beta C \Omega_\beta \otimes J_\beta D \Omega_\beta \Bigr),}
where $C \in {\cal R}_\beta (\O_2)$ and $D \in {\cal R}_\beta (\hat{\O})'$. 
By definition,
$J_\Lambda^2 = \1 $ and $J_\Lambda \Omega_\Lambda = \Omega_\Lambda$.  
Moreover,
$\Omega_\Lambda = V_\Lambda^* (\Omega_\beta \otimes \Omega_\beta)$ 
is the unique --- up to a phase ---
normalized eigenvector for the simple eigenvalue $\{ 0 \}$ of the generator of the
group $t \mapsto \Delta_\Lambda^{it}$.
\vskip .5cm
\noindent
viii.) Assume the maps $\Theta^\pm_{\lambda} \colon {\cal R}_\beta (\O_1) \to \H_\beta $  
\# {  A   \mapsto {\rm e}^{\mp \lambda H_\beta } P^\pm A \Omega_\beta ,
\qquad \lambda > 0 ,} 
where $P^\pm$ denote the projections 
onto the {\sl strictly} positive resp.\ negative spectrum of $H_\beta$, 
are nuclear, and that there exist positive constants
$C (\O_1)$ and $m$ such that the nuclear norm $\| \, . \, \|_1$ 
of $ \Theta^\pm_{\lambda} $ satisfies
\# { \|   \Theta^\pm_{\lambda}\|_1 
\le C (\O_1) \cdot \lambda^{-m}  .}
The split property for the net 
$\O \to {\cal R}_\beta (\O)$ is a consequence of this nuclearity condition [J\"a d]. 
Now let $Q^\pm$ denote the projections 
onto the {\sl strictly} positive resp.\ negative spectrum of $K_\beta$.
It follows that
the maps  $\theta^\pm_\lambda \colon {\cal M}_\beta (\O_1)  \to \H_\beta$,
\# { N  \mapsto {\rm e}^{ \mp \lambda K_\beta } Q^\pm N \Omega_\Lambda,  \qquad \lambda > 0, }
are nuclear. This can be seen as follows:
Let $A \in {\cal R}_\beta (\O_1)$ and $B \in j \bigl( {\cal R}_\beta (\O_1) \bigr)$. 
By definition,
\# { \Delta_\Lambda^\lambda AB \Omega_\Lambda   
=  V^*  \Bigl( \Delta_\beta^{ \lambda} A \Omega_\beta  
\otimes \Delta_\beta^{- \lambda} B \Omega_\beta \Bigr) .}
The maps
$ A \mapsto \Delta^{\pm \lambda}_\beta P^\pm A \Omega_\beta$,  $A \in {\cal R}_\beta (\O_1)$,
and
$B \mapsto \Delta^{\mp \lambda}_\beta P^\mp B \Omega_\beta$,
$B \in j_\beta \bigl( {\cal R}_\beta (\O_1) \bigr)$,
are nuclear for $\lambda > 0$. The tensor product of two nuclear maps is again a
nuclear map and the norm is bounded by the product of the nuclear norms [P].
Furthermore,
\& { Q^+ &= V^* \bigl( P^+ \otimes | \Omega_\beta \ket \bra \Omega_\beta | \bigr) V + 
V^* \bigl( | \Omega_\beta \ket \bra \Omega_\beta | \otimes P^+ \bigr) V
\cr
& \qquad +
V^* \bigl( P^+ \otimes P^+ \bigr) V + V^* \bigl( P^- \otimes P^- \bigr) V}
and
\& { Q^- &= V^* \bigl( P^- \otimes | \Omega_\beta \ket \bra \Omega_\beta | \bigr) V + 
V^* \bigl( | \Omega_\beta \ket \bra \Omega_\beta | \otimes P^- \bigr) V
\cr
& \qquad +
V^* \bigl( P^+ \otimes P^- \bigr) V + V^* \bigl( P^- \otimes P^+ \bigr) V .}
For $\lambda$ large,
the leading contributions to the nuclear norm of the map given
in (77) come from the terms involving the projection 
$| \Omega_\beta \ket \bra \Omega_\beta | $ onto the simple eigenvalue $\{0 \}$ 
on one side of the tensor product. 
Thus
\# { \|  \theta^\pm_{\lambda}  \|   \le 
2 C (\O_1) \cdot  \lambda^{- m} + O \bigl(\lambda^{- 2m} \bigr)  ,}
where $C (\O_1)$ is the constant appearing in the 
nuclearity condition (76) for ${\cal R}_\beta (\O_1)$.
\vskip .5cm
\noindent
ix.) The analyticity properties required in condition (ii) of Theorem 2.2 are satisfied:
For each $N \in {\cal N}:= {\cal M}_\beta (\O)$ the vector valued function
\# { t \mapsto \Delta_\Lambda^{it} N \Omega_\Lambda }
admits an analytic continuation into the strip 
$ {\cal S} (0, 1/2) = \{ z \in \C :
0 \le \Im z \le 1 /2 \} $. 
The boundary values are continuous for $\Im z \nearrow 1/2$ and, since
\& {  J_\Lambda \Delta_\Lambda^{1/2} CD \Omega_\Lambda 
& =  V_\Lambda^*  \Bigl(  
C^* \Omega_\beta \otimes  D^* \Omega_\beta \Bigr)
\cr
& = C^* D^* \Omega_\Lambda
= (CD)^* \Omega_\Lambda }
for all  $C \in {\cal R}_\beta (\O)$ and 
$D \in j_\beta \bigl( {\cal R}_\beta (\O) \bigr)$,  they satisfy
\# { J_\Lambda \Delta_\Lambda^{1/2} N \Psi = N^* \Psi.}
Furthermore,
\# { J_\Lambda \Delta_\Lambda^{iz} N \Omega_\Lambda = \Delta_\Lambda^{i \bar z} 
J_\Lambda N \Omega_\Lambda \qquad \forall z \in {\cal S} ( 0, 1/2),
\quad N \in {\cal M}_\beta (\O). }
Note that $J_\Lambda$ and $\Delta_\Lambda$ are {\sl not}
the modular objects associated to the pair 
$\bigl( {\cal M}_\beta (\O), \Omega_\Lambda \bigr)$. 
\vskip .5cm
\noindent
x.) By definition (73)
\# {  \Delta_\Lambda^{it} \,  \M_\beta (\O_1) \Delta_\Lambda^{-it} \subset \M_\beta 
(\O_1 + t \beta \cdot e)  \qquad \hbox{for} \quad |t| < \delta.}
Inspecting the definition (74) of $J_\Lambda$ 
we find
\# { J_\Lambda \, {\cal M}_\beta (\hat{\O})' J_\Lambda = V_\Lambda^* \bigl( \1 \otimes J_\beta
{\cal M}_\beta (\hat {\O})' J_\beta \bigr) V_\Lambda = {\cal M}_\beta (\hat {\O})'.}
Therefore
\# { \Bigl[ \Delta_\Lambda^{it} \,  \M_\beta (\O_1) 
\Delta_\Lambda^{-it} \, , \, 
\M_\beta (\hat{\O})' \vee j_\Lambda \bigl( \M_\beta (\hat{\O})'\bigr)
\Bigr] = 0 \qquad \hbox{for} \quad |t| < \delta }
reduces to
\# { \Bigl[ \Delta_\Lambda^{it} \,   \M_\beta (\O_1)   
\Delta_\Lambda^{-it} \, , \, 
\M_\beta (\hat{\O})' \Bigr] = 0 \qquad \hbox{for} \quad |t| < \delta,}
which is a direct consequence of (86). 
\vskip .5cm
\noindent
xi.) The product vector $\Omega_\Gamma$ satisfies 
\# {  ( \Omega_\Gamma \, , \, SR \Omega_\Gamma)  = ( \Omega_\beta \, , \, S \Omega_\beta)
( \Omega_\beta \, , \, R \Omega_\beta)
\qquad \forall S \in {\cal S}_\Gamma, \quad R \in {\cal S}_\Gamma' .}
Consequently, there exists an isometry $W_\Gamma$ such that
\# {W_\Gamma R \Omega_\beta = R \Omega_\Gamma \qquad \forall R \in {\cal S}_\Gamma'.}
Inspecting (91), we notice that $W_\Gamma \in {\cal S}_\Gamma$.
Moreover,
\# {  ( \Omega_\Gamma \, , \, T \Omega_\Gamma)  = ( \Omega_\Lambda \, , \, T \Omega_\Lambda)
\qquad \forall T \in {\cal S}_\Gamma \vee {\cal R}_\beta (\hat{\O})'. }
If $M \in {\cal M}_\beta (\hat{\O})'$ and 
$N \in {\cal M}_\beta (\O)$, then $MN \in S_\Gamma \vee {\cal R}_\beta (\hat{\O})'$.
Hence
\& { (\Omega_\beta \, , \, M N \Omega_\beta) & - ( \Omega_\beta \, , \, M \Omega_\beta )
( \Omega_\beta \, , \, N \Omega_\beta ) = 
\cr
& = (\Omega_\Gamma \, , \, M W_\Gamma N W^*_\Gamma \Omega_\Gamma) - 
( \Omega_\Lambda \, , \, M \Omega_\Lambda ) ( \Omega_\Gamma \, , \, 
W_\Gamma N W^*_\Gamma \Omega_\Gamma )
\cr
& = (\Omega_\Lambda \, , \, M W_\Gamma N W^*_\Gamma \Omega_\Lambda) - 
( \Omega_\Lambda \, , \, M \Omega_\Lambda ) 
(\Omega_\Lambda \, , \, W_\Gamma NW^*_\Gamma  \Omega_\Lambda).}
We note that $W_\Gamma N W^*_\Gamma \subset S_\Gamma \vee 
{\cal M}_\beta (\O) \subset {\cal M}_\beta (\O_1)$.

\vskip .8cm
These observations are summarized as follows:

\Cor{Consider a quadruple of space-time regions $\Lambda := (\O, \O_1, \O_2, \hat{\O})$ 
such that
\# { \O \subset \subset \O_1 + te \subset \O_2 \subset \subset \hat{\O} \qquad 
\forall |t| < \delta, 
\qquad \delta > 0 .}
Assume the maps $\Theta^\pm_{\lambda} \colon {\cal R}_\beta (\O_1) \to \H_\beta $  
\# {  A   \mapsto {\rm e}^{\mp \lambda H_\beta } P^\pm A \Omega_\beta ,
\qquad \lambda > 0 ,} 
where $P^\pm$ denote the projections 
onto the {\sl strictly} positive resp.\ negative spectrum of $H_\beta$, 
are nuclear and that there exist positive constants
$C (\O_1)$ and $m$ such that the nuclear norm $\| \, . \, \|_1$ 
of $ \Theta^\pm_{\lambda} $ satisfies
\# { \|   \Theta^\pm_{\lambda}\|_1 
\le C (\O_1) \cdot \lambda^{-m}  .}
\vskip .3cm
\noindent
It follows that for two arbitrary families of operators $N_j \in {\cal M}_\beta (\O)$ and
$M_j \in {\cal M}_\beta (\hat{\O})'$, $j \in \{  1 , \ldots , j_\circ \}$  
\& {\Bigl| \Bigl( \Omega_\beta \, , \, \sum_{j=1}^{j_\circ} M_j N_j \Omega_\beta \Bigr)
& -  \sum_{j=1}^{j_\circ} \bigl( \Omega_\beta \, , \, M_j \Omega_\beta \bigr) 
\bigl( \Omega_\beta \, , \, N_j \Omega_\beta \bigr) \Bigr|  
\cr
&
\le
\bigl( 8 {\rm e}^m  + 2 \bigr) C 
\cdot n_\circ^{-m}
\Bigl\| \sum_{j=1}^{j_\circ} M_j N_j \Bigr\| + O \bigl( n_\circ^{-2m} \bigr) , }
where $n_\circ := \inf  \{ n \in \N : \delta^{m \over m+1} \le n \le \delta \}$. }

\vskip 1cm

\Hl{The Convergence of the Universal Localizing Map}

\noindent
In the vacuum sector, the split property
allows us to approximate global quantities (e.g., the generators of symmetry transformations) 
by local ones~[BDL]. As the space-time regions tend to $\R^4$, the local quantities converge 
to the global ones [D'ADFL].  

The situation in the thermal case is slightly more involved.
Set $\Upsilon_i :=(\O_i, \hat{\O}_i)$ for $\O_i \subset
\subset \hat {\O}_i$ and $i \in \N$. Assume $\O_i, \hat {\O}_i$ tend to  $\R^4 $ 
as $i \to \infty$. If the sequence of split inclusion 
\# {{\cal R}_\beta (\O_i) \subset {\cal S}_{\Upsilon_i}  
\subset {\cal R}_\beta (\hat {\O}_i) }
is used to define a sequence of `universal localizing maps', then this sequence
can not converge against the identity as $i \to \infty$
(see [D'ADFL]). 
Therefore we have to look out for different split inclusions. The tensor product of 
two type I factors is again of type I [S 71, Prop.\ 2.6.2]. 
It follows that there exist type I factors, namely
${\cal K}_{\Upsilon_i} := {\cal S}_{\Upsilon_i} \vee J_\beta {\cal S}_{\Upsilon_i} J_\beta$, 
such that 
\# {{\cal M}_\beta (\O_i) \subset {\cal K}_{\Upsilon_i}  
\subset {\cal M}_\beta (\hat {\O}_i).}
This split inclusion is standard\footnote{$^\star$}{\sevenrm
A split inclusion $\scriptstyle {\cal A} \subset {\cal B}$ is called standard, if there exists
a vector $\scriptstyle \Omega$ which is cyclic for $\scriptstyle {\cal A}' \wedge {\cal B}$ 
as well as for $\scriptstyle {\cal A}$ and $\scriptstyle {\cal B}$.} [DL]. Consequently,
there exists a unique vector $\eta_{\Upsilon_i}$ in the natural positive cone
${\cal P}^\natural \bigl({\cal M}_\beta (\O_i) \vee {\cal M}_\beta (\hat {\O}_i)', 
\Omega_\beta \bigr)$ such that
\# {( \eta_{\Upsilon_i} \, , \, M_i N_i \eta_{\Upsilon_i}) 
= (\Omega_\beta \, , \, M_i \Omega_\beta ) 
(\Omega_\beta \, , \, N_i \Omega_\beta )}
for all $ N_i \in 
{\cal M}_\beta (\O_i)$ and all $ M_i \in {\cal M}_\beta (\hat {\O}_i)'$. 
The product vectors $\eta_{\Upsilon_i}$ can now be utilized in the following

\Def{Consider two open and bounded space-time regions $\O$ and $\hat{\O}$ such that the 
closure of $\O$ is contained in the interior of $\hat{\O}$.
The universal localizing map $\psi_\Upsilon \colon \B (\H_\beta) \to {\cal K}_\Upsilon$ 
associated with the pair $\Upsilon:=
(\O, \hat {\O})$ is given by
\# { T \mapsto  U_\Upsilon^{-1} \bigl( T \otimes \1 \bigr) U_\Upsilon,
\qquad T \in \B (\H_\beta),}
where 
$U_\Upsilon \colon  \H_\beta \to \H_\beta \otimes \H_\beta$ is given by
linear extension of
\# {U_\Upsilon MN \eta_\Upsilon := M \Omega_\beta \otimes 
N \Omega_\beta }
for all $ N \in 
{\cal M}_\beta (\O)$ and all $ M \in {\cal M}_\beta (\hat {\O})'$. 
} 

The map $\psi_\Upsilon$ is a *-isomorphism of $\B(\H_\beta)$ onto the canonical type I factor
${\cal K}_\Upsilon$ between 
${\cal M}_\beta (\O)$ and ${\cal M}_\beta (\hat {\O})$
and it acts trivial on ${\cal M}_\beta (\O)$.

\Prop{Consider a sequence of pairs $\Upsilon_i := (\O_i, \hat {\O}_i)$ of space-time regions
with diameters $r_i$ and $\hat{r}_i$, respectively, such that
\# { \hat{r}_i = r_i^\gamma \qquad \hbox{and} 
\qquad \gamma > {d( m+1) \over m^2} .}
Assume the maps $\vartheta_{\lambda, {\cal O}_i}
\colon {\cal R}_\beta (\O_i) \to \H_\beta $ 
\# {  A   \mapsto \Delta^{\pm \lambda} P^\pm  A \Omega_\beta , \qquad \lambda > 0,} 
are nuclear and satisfies
\# { \|  \vartheta_{\lambda, {\cal O}_i}  \|_1 
\le const \cdot r_i^d  \lambda^{-m}  . }
It follows that $\Psi_{\Upsilon_i}$ converges 
pointwise strongly to the identity, i.e.,
\# { s-\lim_{i \to \infty} \Psi_{\Upsilon_i} (T) = T \qquad 
\forall T \in \B(\H_\beta).}
}

\Pr{The norm distance between $\eta_{\Upsilon_i} \in {\cal P}^\natural
\bigl({\cal M} (\O_i) \vee {\cal M} (\hat{\O}_i)', \Omega_\beta \bigr)$  and $\Omega_\beta$
is bounded by
\# { \| \eta_{\Upsilon_i} - \Omega_\beta  \|^2 \le \sup_{\| M_i \| = 1} \bigr| 
(\eta_{\Upsilon_i} \, , \, M_i \eta_{\Upsilon_i} ) -
(\Omega_\beta \, , \, M_i \Omega_\beta ) \bigr|,
}
where the supremum has to be taken w.r.t.\ $M_i \in {\cal M} (\O_i) \vee {\cal M} (\hat{\O}_i)'$
[BR, 2.5.31].
Consequently, the convergence of the universal localizing maps follows from 
Corollary 3.3: For $\delta_i := \hat{r}_i - r_i$ sufficiently large
\# { \| \eta_{\Upsilon_i} - \Omega_\beta  \|^2 \le const' \cdot r_i^d   
\delta_i^{-{ m^2 \over m +1}}.}
By assumption
\# {\hat{r}_i = r_i^\gamma \qquad \hbox{with } \qquad \gamma > {d( m+1) \over m^2}.}
It follows that the r.h.s.\ of (108) tends to zero as $i \to \infty$. Moreover, for 
$T \in \B(\H_\beta)$ and 
$N \in {\cal M}_\beta (\O)$ we find
\# { U^{-1}_{\Upsilon_i} \bigl( TN \Omega_\beta \otimes \Omega_\beta)
\to TN \Omega_\beta \qquad \hbox{as} \quad  i \to \infty .}
Applying the definition of $\psi_{\Upsilon_i}$ we conclude that
\# { \psi_{\Upsilon_i} (T)   N \eta_{\Upsilon_i}
\to TN \Omega_\beta  \qquad \hbox{as} \quad  i \to \infty .}
Now $\eta_{\Upsilon_i}$ tends to $\Omega_\beta$, 
$\Omega_\beta$ is cyclic for ${\cal M}_\beta (\O)$
and the maps $\psi_{\Upsilon_i}$, $i \in \N$, are uniformly bounded, 
hence $\psi_{\Upsilon_i} (T)$ converges
strongly to~$T$  as $i \to \infty$ [D'ADFL].}

\Hl{Appendix: Wedge-shaped Regions}

\noindent
We consider a local quantum field theory in the vacuum representation $\pi$,
specified by a net of von Neumann algebras
$ \O \to {\cal R} (\O) := \pi \bigl( \A(\O) \bigr)''$.  
Let
\# {W := \{ x \in \R^4 : |x_0| < x_3 \} }
denote a wedge-shaped region in Minkowski space. We assume that the modular operator $\Delta_W$
associated with the algebra ${\cal R} (W)$ and the vacuum vector $\Omega$ 
implements the Lorentz boosts, i.e.,
\# {\Delta_W^{it} {\cal R} (\O) \Delta_W^{-it} = {\cal R} \bigl( M( \pi t) \O \bigr) ,
\qquad \O \subset W,}
where $M(t)$ denotes the Lorentz matrix
\# { M(t) :=\pmatrix{\cosh (t) & 0 &0 & \sinh (t) \cr
                     0         & 1 &0 &         0 \cr
                     0         & 0 &1 &         0 \cr
                     \sinh (t) & 0 &0 & \cosh (t) \cr}.}
Bisognano and Wichmann [BW a,b] have shown that (113) is generically the case, if the 
net of local observables is constructed from a Wightman field theory.  
Moreover,
Borchers~[Bo] has shown that in 1+1 space-time dimensions equation (113) is a consequence of
the standard assumptions of algebraic quantum field theory. 
The algebra ${\cal R} (W)$ associated with a wedge is of type $III_1$ [Dr],
thus the spectrum of $\log \Delta_W$ has a (up to a phase) unique
eigenvector for the simple eigenvalue $\{ 0\}$, 
namely the vacuum vector $\Omega$,  and
otherwise continuous spectrum on the whole real axis.

The conditions (i-iii) of Theorem 2.1 are fulfilled if we 
consider two algebras ${\cal R} (\O_1) \subset {\cal R} (W)$
and ${\cal R} (\O_2) \subset {\cal R}:= \pi (\A)''$
such that
\# { \Delta_W^{it} {\cal R} (\O_1)  \Delta_W^{-it}  
\subset {\cal R} (\O_2)'  , \qquad  |t| < \delta, \quad \delta > 0,}
and put 
\# { \matrix{ {\cal N} &:= & {\cal R} (\O_1) ,  & \qquad 
{\cal M} & :=  & {\cal R} (\O_2)  ,
&\qquad \Psi &:= &\Omega ,
\cr
&&& &&&\cr
                 U (t) &:= & \Delta_W^{it}, & \qquad  I & :=  & J_W . 
&& \cr}}
Spectral information on the vacuum Hamilton operator ${\bf H}$ can be 
used to estimate
\# { \| {\rm e}^{ \mp \lambda H} P^\pm N \Psi \| , 
\qquad N \in {\cal R} (\O_1).  }
One can proceed along the lines discussed in [BD'AL], see for example, the last
inequality on p.\ 125. But
since we do not expect any improvement of the existing vacuum cluster theorems,
we do not dwell on this point.

\vskip 1cm
\noindent
{\it  Acknowledgements.\/}
\noindent
Kind hospitality of the Erwin Schr\"odinger Institut (ESI) Vienna and the
Dipartimento di Matematica, Universit\'a di Roma ``Tor Vergata'' is 
gratefully acknowleged. This work was financed by
the Fond zur F\"orderung der Wissenschaft\-lichen 
Forschung in Austria, Proj.\ Nr.\ P10629 PHY and a Fellowship
of the Operator Algebras Network, EC TMR-Program. 

\vskip 1cm

\noindent
{\fourteenrm References}
\nobreak
\vskip .3cm
\nobreak
\halign{   &  \vtop { \parindent=0pt \hsize=33em
                            \strut  # \strut} \cr 
\REF
{AHR}
{Araki, H., Hepp, K., Ruelle, D.}  
                          {On the asymptotic behavior of Wightman functions
                           in space-like directions}
                          {Helv.\ Phys.\ Acta}
                          {35}     {164--174}
                          {1962}
\REF
{Bo}
{Borchers, H.J.}     
                                   {The CPT-theorem in two-dimensional theories of local
                                    observables}
                                   {\CMP}
                                   {143}    {315--332}
                                   {1992}
\REF
{BDL}
{Buchholz, D., Doplicher, S.\ and Longo, R.}
                                    {On Noether's theorem in quantum field theory}
                                    {Ann.\ Phys.}
                                    {170}  {1--17} 
                                    {1986}
\BOOK
{BR}  
{Bratteli, O.\ and Robinson, D.W.} {Operator Algebras and Quantum Statistical Mechanics~I,II} 
                                  {Sprin\-ger-Verlag, New York-Heidelberg-Berlin} 
                                  {1981}
\REF
{BW a}
{Bisognano, J.\ and Wichmann, E.H.}  
                          {On the duality condition for a Hermitian scalar field}
                          {\JMP}
                          {16}     {985--1007}
                          {1975}
\REF
{BW b}
{Bisognano, J.\ and Wichmann, E.H.}  
                          {On the duality condition for quantum fields}
                          {\JMP}
                          {17}     {303--321}
                          {1976}
\HEP
{D}
{Davidson, D.R.}     
                                   {Cluster estimates and analytic wavefunctions}
                                   {hep-th/9406104}
\REF
{Dr}
{Driessler, W.}     
                                   {Comments on lightlike translations and applications in 
                                    relativistic quantum field theory}
                                   {\CMP}
                                   {44}    {133--141}
                                   {1975}
{\sevenbf [D'ADFL]}  & \hskip -9.5cm \vtop {
                    {\sevenrm D'Antoni, C., Doplicher, S., Fredenhagen, K., and Longo, R.,} 
                    {\sevensl Convergence of local charges and continuity properties of 
                        $\scriptstyle W^*$-inclusions,} 
                                {\sevenrm \CMP} 
                                {\sevenbf 110,} 
                                {\sevenrm 321--344} 
                                {\sevenrm (1987)}
                                {\sevensl and} 
                                {\sevenrm Erratum} 
                                {\sevenrm \CMP} 
                                {\sevenbf 116,} 
                                {\sevenrm 321--344}
                                {\sevenrm (1988)} } \cr
\REF
{DL}
{Doplicher, S.\ and Longo, R.}      {Standard and split inclusions of von Neumann algebras}
                                   {Invent.\ Math.}
                                   {73}    {493--536}
                                   {1984}
\REF
{F}
{Fredenhagen, K.}      {A remark on the cluster theorem}
                       {\CMP}
                       {97} {461--463}
                       {1985}
\BOOK
{H}
{Haag, R.}    {Local Quantum Physics: Fields, Particles, Algebras} 
              {Springer-Verlag, Berlin-Heidelberg-New York} 
              {1992}
\REF
{HHW}
{Haag, R., Hugenholtz, N.M.\ and Winnink, M.}
                          {On the equilibrium states in quantum statistical mechanics}  
                          {\CMP}	
                          {5}	{215--236}
                          {1967}
\REF
{J\"a a}
{J\"akel, C.D.}                   {Decay of spatial correlations in thermal states}
                                  {Ann.\ l'Inst.\ H.\ Poincar\'e} 
                          {69}	{425--440}
                          {1998}
\HEP
{J\"a b}
{J\"akel, C.D.}                   {On the relation between KMS states for different temperatures}
                                  {hep-th/9803245} 
\HEP
{J\"a c}
{J\"akel, C.D.}                   {Two algebraic properties of thermal quantum field theories}
                                  {hep-th/9901015} 
\HEP
{J\"a d}
{J\"akel, C.D.}                   {Nuclearity and split for thermal quantum field theories}
                                  {hep-th/9811227} 
\HEP
{J\"a e}
{J\"akel, C.D.}                   {The Reeh-Schlieder property for thermal field theories}
                                  {hep-th/9904049} 
\BOOK
{KR}
{Kadison, R.V.\ and Ringrose, J.R.}     {Fundamentals of the theory of operator algebras II} 
                   {Academic Press, New York}  
                  {1986}
\BOOK
{P}
{Pietsch, A.}     {Nuclear locally convex spaces} 
                  {Springer-Verlag, Berlin-Heidelberg-New York}  
                  {1972}
\BOOK
{Sa 71}
{Sakai, S.}     {$\scriptstyle C^*$-Algebras and $\scriptstyle W^*$-Algebras} 
                     {Springer-Verlag, Berlin-Heidel\-berg-New York}  
                     {1971}
\BOOK
{St}
{Str\v atila, S.}   {Modular Theory in Operator Algebras}
                 {Abacus Press} 
                 {1981}
\BOOK
{SW}
{Streater, R.F.\ and Wightman, A.S.}   {PCT, Spin and Statistics and all that}
                 {Benjamin, New York} 
                 {1964}
\BOOK
{Ta}
{Takesaki, M.}   {Tomita's Theory of Modular Hilbert Algebras and its Application}
                 {Springer, Lecture Notes in Mathematics, Berlin} 
                 {1970}
\REF
{TW}
{Thomas, L.J.\ and Wichman, E.H.}      {About matrix elements of spatial 
                                        translations in local field theories}
                                   {\LMP} 
                                   {28} {49--58}
						                             {1993}
\cr}

\bye